\documentclass[twocolumn,english,prl]{revtex4}
\usepackage[latin9]{inputenc}
\pdfoutput=1
\usepackage{amsmath}
\usepackage{amssymb}
\usepackage{graphicx}
\usepackage{esint}

\makeatletter
\@ifundefined{textcolor}{}
{%
 \definecolor{BLACK}{gray}{0}
 \definecolor{WHITE}{gray}{1}
 \definecolor{RED}{rgb}{1,0,0}
 \definecolor{GREEN}{rgb}{0,1,0}
 \definecolor{BLUE}{rgb}{0,0,1}
 \definecolor{CYAN}{cmyk}{1,0,0,0}
 \definecolor{MAGENTA}{cmyk}{0,1,0,0}
 \definecolor{YELLOW}{cmyk}{0,0,1,0}
}

\usepackage{upgreek}

\renewcommand{\vec}[1]{\mathbf{#1}}

\makeatother

\usepackage{babel}
\begin{document}

\title{Dirac Phase interferometer in a plasmonic waveguide}

\author{V. Savinov}

\email{vs1106@orc.soton.ac.uk}

\selectlanguage{english}%

\affiliation{Optoelectronics Research Centre \& Centre for Photonic Metamaterials,
University of Southampton SO17 1BJ, UK}

\author{N. I. Zheludev}

\affiliation{Optoelectronics Research Centre \& Centre for Photonic Metamaterials,
University of Southampton SO17 1BJ, UK}

\affiliation{Centre for Disruptive Photonic Technologies, TPI, Nanyang Technological
University, Singapore 637371, Singapore}
\begin{abstract}
By viewing plasmon waves in metallic waveguides as propagating electric
and magnetic dipoles we show that according to laws of quantum mechanics
they will acquire additional phase when propagating through space
with static magnetic field. The new effect is physically different
from conventional magneto-plasmonic phenomena and is sufficiently
strong to observe it under routinely accessible experimental conditions. 
\end{abstract}
\maketitle
% Intro. First para = the weird AB

The Aharonov-Bohm (AB) effect is both one of the most celebrated and
vigorously debated effects in quantum physics \cite{Siday49,Bohm59}.
It is a quantum mechanical topological phenomenon in which phase of
an electrically charged particle is affected by the presence of magnetic
field, even if the particle is confined to a region in which the field
is zero. The modern interpretation of this phenomenon is that the
dynamics of the electrons are affected by the magnetic vector potential
rather than by the magnetic field. 

% Next the AC and HMW effects also introduce the Dirac Phase

The Aharonov-Bohm effect arises as a result of Dirac's magnetic phase
factor \cite{Dirac31,Berry80,Berry10}, the extra phase gained by
the electrons propagating in the electromagnetic potential. There
are other similar effects that also arise due to Dirac's phase factor.
In particular, in the Aharonov-Casher (AC) effect the phase is gained
by the particle with permanent magnetic dipole due to propagation
in electric field \cite{AC84}, whilst in the He-McKellar-Wilkens
(HMW) effect \cite{HeMcKellar93,Wilkens94}, the phase is gained by
the particles with permanent electric dipole that propagate in magnetic
field.

% Now the experimental tests of all three effects. No focus on JJ as I am currently not relying 
% on this in the main text.

All three effects described above have now been experimentally observed.
The Aharonov-Bohm effect was demonstrated by Tonomura \cite{Tonomura86}.
In his experiment electrons were made to pass through and around a
toroidal solenoid and then their interference pattern was recorded.
The Aharonov-Casher effect was tested with neutron \cite{Cimmino89}
and atomic interferometry \cite{Sangster93}, and was also observed
in solid-state systems \cite{Konig06,Qu11} as well as in Josephson
junctions \cite{Elion93,Pop12}. The He-McKellar-Wilkens effect was
only confirmed recently by interfering trains of polarized lithium
atoms \cite{Lepoutre12,Gillot13,LepoutrePRA13_I,LepoutrePRA13_II}.

% Now will have to explain what the He-McKellar-Wilkens effect is in detail

\begin{figure}
\includegraphics{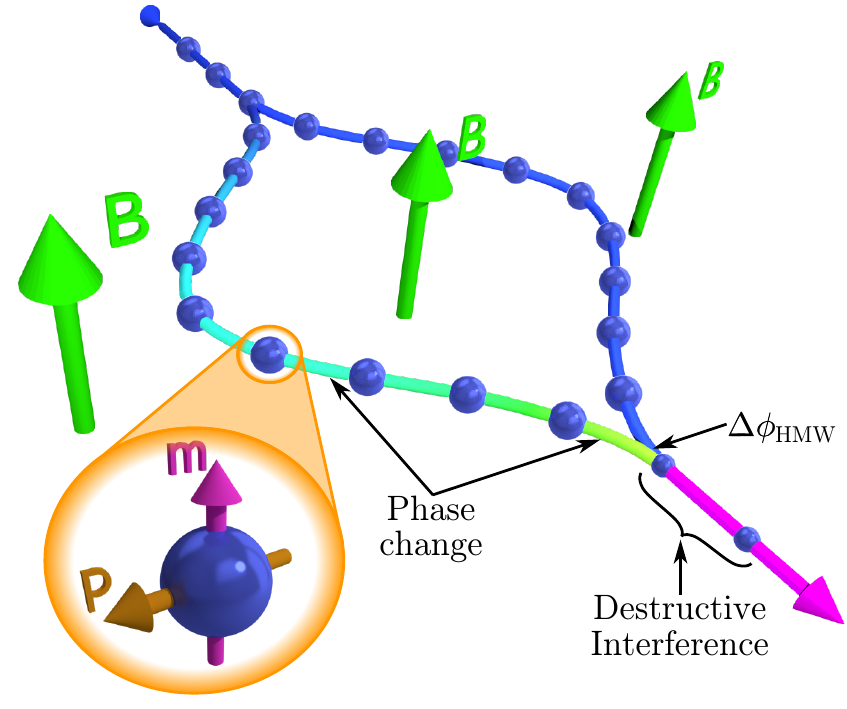}

\protect\caption{\textbf{Illustration of the He-McKellar-Wilkens effect.} Neutral particles
with permanent electric ($\vec{p}$) and magnetic ($\vec{m}$) dipoles
are passed through an interferometer in the presence of externally
applied magnetic field ($\vec{B}$). Propagation of particles leads
to change in their phase. If the magnetic field is different at the
two arms of the interferometer, or if the lengths of the two arms
are different, the phase difference ($\Delta\phi_{HMW}$) can be detected
through interference of the particles.}

\label{fig:HMWEff}
\end{figure}

The focus of our paper will be the He-McKellar-Wilkens effect illustrated
in Fig.~\ref{fig:HMWEff}. Consider a train of traveling particles
with permanent electric ($\vec{p}$) and magnetic dipole ($\vec{m}$)
moments. The train is passed through a particle interferometer where
it is split in two branches which are routed along different paths,
and are then joined back together into a single train. If the interferometer
is placed into magnetic field ($\vec{B}$), the phase of the particles
in the two branches of the train will be shifted by an amount that
will depend on the magnetic field strength. This will lead to magnetically
tunable particle interference at the output of the interferometer. 

% Waisted enough time! What will we do? Quickly explain the core idea and the main assumtpion of the method
% Then start moving towards establishing the framework to compute the magnitude of the effect 
% i.e. the nitty-gritty

In this paper we will show that the He-McKellar-Wilkens effect can
be observed with surface plasmon-polaritons propagating in static
magnetic field. Surface plasmon-polaritons are coupled oscillations
of light and electron plasma in metals. These propagating coupled
oscillations can be described as quantum-mechanical particles propagating
along the waveguide trajectory. The correspondence between the wave
and particle pictures is established by assuming that particles carry
electric and magnetic dipole moments corresponding to the classical
polarization ($\vec{P}$) and magnetization ($\vec{M}$) of the guided
wave. We will now show that the phase gain by plasmon-polaritons due
to He-McKellar-Wilkens effect will be observable under routinely accessible
experimental conditions in a plasmonic waveguide. 

The Lagrangian for a neutral particle with permanent electric ($\vec{p}$)
and magnetic ($\vec{m}$) dipole moments moving at velocity $\vec{v}$
through electromagnetic field is \cite{AC84,Wilkens94,Anandan00,Gillot13}
:
\begin{multline}
L=L_{kin}+L_{self}+\\
+\vec{E}.\left(\vec{p}+\frac{1}{c^{2}}\vec{v}\times\vec{m}\right)+\vec{B}.\left(\vec{m}-\vec{v}\times\vec{p}\right)\label{eq:MainLagrangian}
\end{multline}

Where $\vec{E}$ and $\vec{B}$ denote the electric and magnetic fields,
respectively. Above, $L_{kin}$ denotes the kinetic part of the Lagrangian
and $L_{self}$ denotes the Lagrangian due to interaction of the dipoles
with their fields. These two terms will be of no interest to us in
what is to follow, as the effect in question is related to the last
two terms in the Lagrangian. 

% Specialize in the HMW effect

A quantum mechanical particle described by the Lagrangian from Eq.~(\ref{eq:MainLagrangian})
will gain phase as a result of propagation. The path-specific expression
for the phase gain ($\Delta\phi$) due to transition from point $\vec{r}_{a}$
at time $t_{a}$ to point $\vec{r}_{b}$ at time $t_{b}$ can be written
in terms of action ($S$) divided by reduced Planck constant ($\hbar$):
\begin{multline}
\Delta\phi=S[\vec{r}]/\hbar=\frac{1}{\hbar}\int_{t_{a}}^{t_{b}}dt\,L=\\
=\Delta\phi_{kin}+\Delta\phi_{self}+\\
+\frac{1}{\hbar}\int_{t_{a}}^{t_{b}}dt\,\left(\vec{p}.\vec{E}+\vec{m}.\vec{B}\right)+\\
+\frac{1}{\hbar}\int_{\vec{r}_{a}}^{\vec{r}_{b}}d\vec{r}.\left(\frac{1}{c^{2}}\vec{m}\times\vec{E}-\vec{p}\times\vec{B}\right)\label{eq:PhaseGain}
\end{multline}

In Eq.~(\ref{eq:PhaseGain}) the fourth term includes phase gained
due to increased or reduced energy of the electric and magnetic dipoles
in the electromagnetic field. The last term of Eq.~(\ref{eq:PhaseGain})
differs significantly from the preceding terms in being expressed
through an integral along a path rather than a temporal integral.
It is relevant to both the He-McKellar-Wilkens and the Aharonov-Casher
effects. For the rest of this paper we will concentrate on the He-McKellar-Wilkens
effect, we therefore set $\vec{E}=\vec{0}$. The phase gained by the
particle as a direct result of propagation in the magnetic field,
without the additional terms, is then given by:
\begin{multline}
\Delta\phi_{HMW}=\Delta\phi-\left(\Delta\phi_{kin}+\Delta\phi_{self}\right)=\\
=\frac{1}{\hbar}\int_{t_{a}}^{t_{b}}dt\,\vec{m}.\vec{B}-\frac{1}{\hbar}\int_{\vec{r}_{a}}^{\vec{r}_{b}}d\vec{r}.\left(\vec{p}\times\vec{B}\right)\label{eq:JustHMW}
\end{multline}

% Move to our idea with the waveguide. motivate the idea and state it

\begin{figure}
\includegraphics{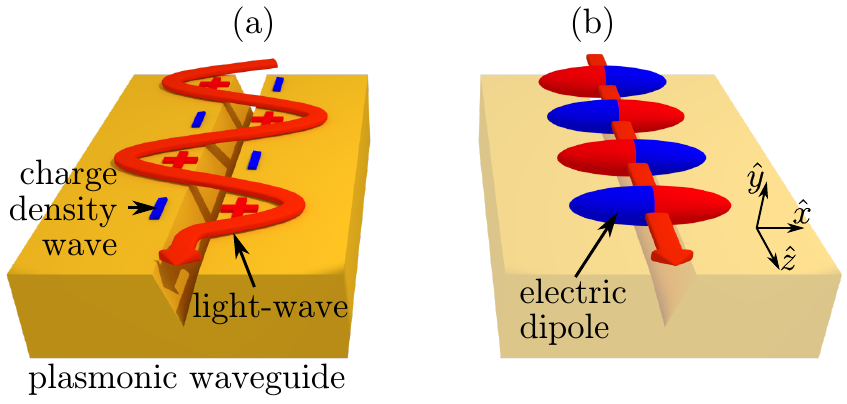}

\protect\caption{\textbf{Viewing plasmonic waveguide with guided light-wave as a stream
of propagating dipoles.} \textbf{(a)}~A sketch of a light-wave propagating
along a plasmonic waveguide. The propagating radiation is accompanied
by induced co-propagating charge density wave (shown with `+' and
`-' signs). \textbf{(b)}~The charge density wave propagating along
the waveguide (as shown in (a)) can be viewed as a stream of anti-aligned
dipoles propagating along the waveguide. For simplicity only the electric
dipoles are shown in the sketch, in reality however there are also
co-propagating magnetic dipoles aligned along the y-axis.}

\label{fig:PlasmonHMWConcept}
\end{figure}

The He-McKellar-Wilkens effect is significantly more difficult to
observe than the Aharonov-Bohm and Aharonov-Casher phenomena. Indeed,
due to lack of elementary particles with electric dipole, the recent
observations of this effect have been obtained by \emph{inducing}
a polarization in otherwise non-polarized charge-current distribution
of the lithium atoms \cite{Lepoutre12,Gillot13,LepoutrePRA13_I,LepoutrePRA13_II}.
We argue that it should be possible to observe the He-McKellar-Wilkens
effect in the traveling waves of \emph{induced polarization} in metals,
i.e. surface plasmon-polaritons. For simplicity we shall consider
plasmon waves in a V-groove plasmonic waveguide, as is shown in Fig.~\ref{fig:PlasmonHMWConcept}.
Since in a linear waveguide the induced polarization scales with amplitude
of the guided mode, one should expect the He-McKellar-Wilkens phase
shift in the waveguide to be a function of the power of the guided
mode. 

% Explain the details of the idea. Including the method for finding the electric and magnetic dipoles
% Set up the waveguide specifics

\begin{figure}
\includegraphics{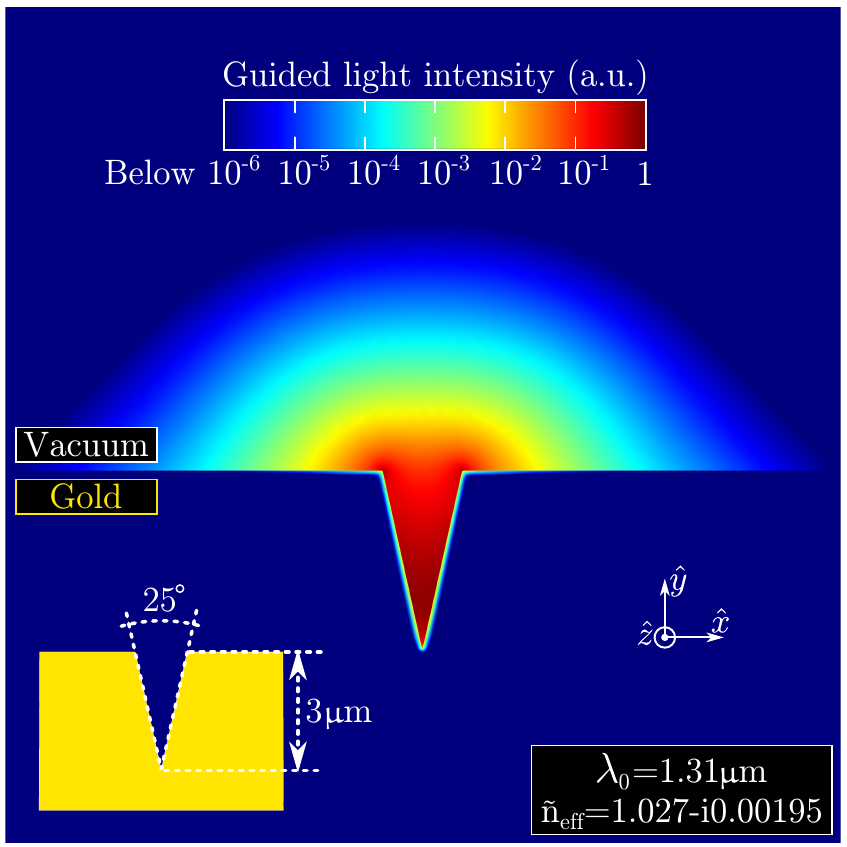}

\protect\caption{\textbf{Simulation of the guided mode in the exemplary waveguide.}
The main plot shows the log-scale colourmap of the intensity of guided
radiation in the waveguide. The waveguide is made of gold and is operating
in free-space environment. The free-space wavelength of the guided
mode is $\lambda_{0}=1.31\,\upmu\mbox{m}$. The inset on the bottom
left-hand side shows the details of the waveguide geometry.}

\label{fig:2DComsol}
\end{figure}

For the following analysis we will adopt a specific waveguide geometry
with opening angle of $25^{\circ}$ and depth of $3\,\upmu\mbox{m}$
(see Fig.~\ref{fig:2DComsol}). We will also assume that the waveguide
is made of gold (dielectric constant taken from Ref.~{[}\citenum{ChristyGold72}{]})
and that the ambient environment is vacuum (or air). Furthermore we
shall assume that the free-space wavelength of the light propagating
along the waveguide is $\lambda_{0}=1.31\,\upmu\mbox{m}$ and that
the power carried by the waveguide is $\mathcal{P}=1\,\upmu\mbox{W}$.
These assumptions are in no way mandatory for observation of the effect,
instead they represent a set of conditions that can be easily accessed
in the experiment.

We proceed by breaking up the continuous traveling charge ($\rho$)
and current ($\vec{J}$) density waves into individual half-periods,
and finding the electric ($\vec{p}$) and magnetic ($\vec{m}$) dipole
moments of each half-period using \cite{jackson}:
\begin{align*}
\vec{p}= & \int d^{3}r\,\rho\vec{r}=\frac{1}{i\omega}\int d^{3}r\,\vec{J}\\
\vec{m}= & \frac{1}{2}\int d^{3}r\,\left[\vec{r}\times\vec{J}\right]
\end{align*}

where the integral is taken over the volume occupied by a single half-period.

Numerical simulation of the mode guided by the waveguide, shown in
Fig.~\ref{fig:2DComsol}, allows to determine the effective refractive
index of the mode $\tilde{n}_{eff}=1.027-i0.00195$ as well as the
distribution of the electric field ($\vec{E}_{mode}(x,y)$) in the
xy-plane, the plane perpendicular to the direction of mode propagation
(along the z-axis). Ignoring the losses in the waveguide ($n_{eff}\equiv\Re(\tilde{n}_{eff})$)
we can write the full distribution of the electric field as $\vec{E}(\vec{r})=\vec{E}_{mode}(x,y)\times\exp\left(-i\frac{2\pi}{\lambda_{0}}n_{eff}z\right)$
and the full current density distribution as $\vec{J}=i\omega\epsilon_{0}\left(\tilde{\epsilon}_{r}^{(Au)}-1\right)\vec{E}$,
where $\epsilon_{0}$ is the free-space permittivity and $\tilde{\epsilon}_{r}^{(Au)}=-79.1-i7.02$
is the dielectric constant of gold at $\lambda_{0}=1.31\,\upmu\mbox{m}$
\cite{ChristyGold72}. Consequently, the expressions for the single-half-period
dipoles become:
\begin{align}
\vec{p}= & \epsilon_{0}\left(\tilde{\epsilon}_{r}^{(Au)}-1\right)\int_{-\frac{\lambda_{0}}{4n_{eff}}}^{\frac{\lambda_{0}}{4n_{eff}}}dz\,\int dxdy\,\vec{E}\left(\vec{r}\right)\label{eq:EDipWaveCrest}\\
\vec{m}= & \frac{i\omega\epsilon_{0}\left(\tilde{\epsilon}_{r}^{(Au)}-1\right)}{2}\int_{-\frac{\lambda_{0}}{4n_{eff}}}^{\frac{\lambda_{0}}{4n_{eff}}}dz\,\int dxdy\,\left[\vec{r}\times\vec{E}\left(\vec{r}\right)\right]\label{eq:MDipWaveCrest}
\end{align}

Numerically evaluating the integrals in Eq.~(\ref{eq:EDipWaveCrest},\ref{eq:MDipWaveCrest})
we find, up to a complex phase, $\vec{p}=\left(9.8\times10^{-27}\,\mbox{C.m}\right)\mathbf{\hat{x}}$
and $\vec{m}=\left(4.1\times10^{-17}\,\mbox{J/T}\right)\mathbf{\hat{y}}$.
One should note that by the nature of the guided mode in the plasmonic
waveguide the two dipoles are perpendicular to each other ($\vec{p}\,\perp\,\vec{m}$),
and to the direction of propagation, at all times ($\vec{p},\,\vec{m}\,\perp\,\vec{v}$,
where $\vec{v}$ is the velocity of the guided mode).

% Explain how we plan to detect the effect. Essentially motivate the \Delta L. The we can find the phase shift
% find the shift for a realistic set of params

\begin{figure}
\includegraphics{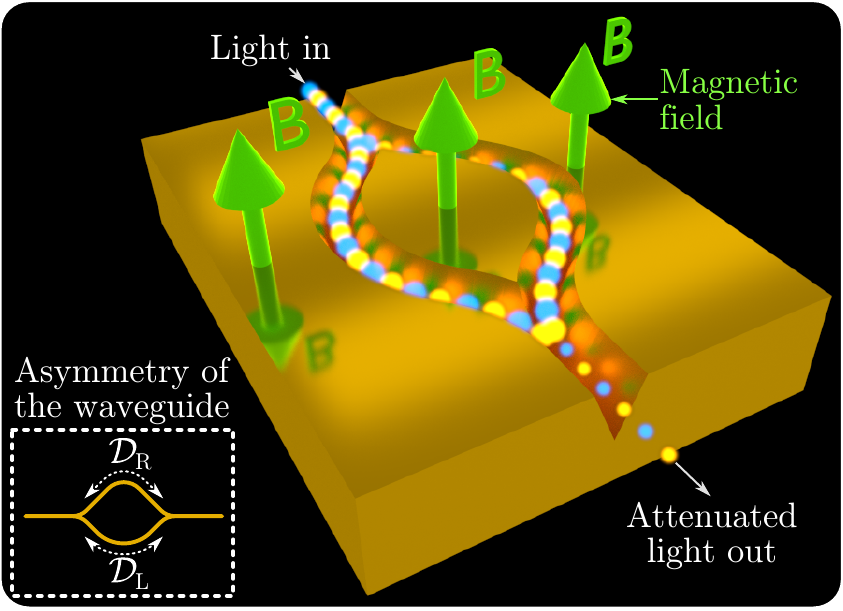}

\protect\caption{\textbf{Detection of He-McKellar-Wilkens effect with a plasmonic V-groove
interferometer.} The guided light-wave and the co-propagating stream
of dipoles (see Fig.~\ref{fig:PlasmonHMWConcept}) is split and recombined
in an interferometer. As is shown in the schematic on the bottom-left
inset, the interferometer is asymmetric with the lengths of the left
($\mathcal{D}_{L}$) and the right ($\mathcal{D}_{R}$) arms being
different ($\Delta\mathcal{D}=\mathcal{D}_{R}-\mathcal{D}_{L}\protect\neq0$).
Applying magnetic field ($\vec{B}$) perpendicular to interferometer
plane will, through the He-McKellar-Wilkens effect, induce an additional
phase difference between surface plasmon-polaritons propagating along
the left and the right arms. The phase difference will be proportional
to the applied magnetic field, and the transmission of the interferometer
will be a function of the applied magnetic field. }

\label{fig:PlasmonInterf}
\end{figure}

The He-McKellar-Wilkens phase shift of the guided plasmonic mode can
be detected in a plasmonic interferometer as is shown in Fig.~\ref{fig:PlasmonInterf}.
Using the fact that the magnetic dipole, electric dipole and the velocity
are always mutually perpendicular ($\vec{p}\:\perp\:\vec{m}\,\perp\:\vec{v}$)
the expressions in Eq.~(\ref{eq:JustHMW}) can be simplified to:
\begin{align}
\Delta\phi_{HMW} & =\frac{B}{\hbar}\left(\mathcal{T}m-\mathcal{D}p\right)\nonumber \\
 & =\frac{B\mathcal{D}}{\hbar}\left(\frac{n_{eff}}{c}m-p\right)\label{eq:PhaseForWG}
\end{align}

Above we have assumed the configuration in which the applied magnetic
field is perpendicular to waveguide plane (as in Fig.~\ref{fig:PlasmonInterf}),
and is therefore parallel to the magnetic dipole ($\vec{B}\,\parallel\,\vec{m}$).
The duration and the length of propagation are denoted with $\mathcal{T}$
and $\mathcal{D}$ respectively. From Eq.~(\ref{eq:PhaseForWG})
it follows that if the interferometer is symmetric the surface plasmon-polaritons
propagating along the different arms will gain the same phase due
to He-McKellar-Wilkens effect. However, an asymmetry of the interferometer
($\Delta\mathcal{D}\neq0$ in Fig.~\ref{fig:PlasmonInterf}) can
give rise to the additional difference in phase of the surface plasmon-polaritons
traveling along the two interferometer arms:
\begin{equation}
\Delta\phi_{HMW}=\frac{B\cdot\Delta\mathcal{D}}{\hbar}\cdot\left(\frac{n_{eff}}{c}m-p\right)\label{eq:pHMW}
\end{equation}

% Substitute. give tha magnitude of the effect

For $\Delta\mathcal{D}=200\,\mbox{nm}$ and $B=10\,\mbox{mT}$ one
finds:
\[
\Delta\phi_{HMW}=\underbrace{2.6\,\mbox{rad}}_{m}-\underbrace{0.19\,\mbox{rad}}_{p}
\]

% Introduce the merit coefficient for the interferometer. Leave the dependence on Delta D. This is ok
% if we are considering a generic interferometer. Since the difference in arms lenght shall, probably, not
% exceed a half-wavelength of the guided wave it is probably fair to leave the coefficient in terms
% of nm.

The phase difference $\Delta\phi_{HMW}$ scales linearly with magnetic
field ($B$), with length difference of the interferometer arms ($\Delta\mathcal{D}$),
and with magnitudes of the two dipole moments ($m$ and $p$). The
dipole magnitudes, in turn, scale as the square-root of the power
($\mathcal{P}$) guided by the plasmonic interferometer. One can therefore
quantify the He-McKellar-Wilkens effect in a plasmonic interferometer
in terms of a dimensional constant:
\[
\Lambda_{HMW}=\frac{\Delta\phi_{HMW}}{\Delta\mathcal{D}\cdot B\cdot\sqrt{\mathcal{P}}}\approx1200\,\mbox{rad/}(\mbox{nm}.\mbox{T}.\sqrt{\mbox{W}})
\]

% Brief discussion. Need to set ourselves apart from the Faraday and co. 
% Our effect is intrinsically nonlinear, nonlocal [may be difficult to prove in detail], quasi material-independent
% Also include NIZ concerns about the scaling of the effect. With power of sqrt(power)?

It is important to note that in plasmonic He-McKellar-Wilkens effect
the transmission of the interferometer depends both on the applied
magnetic field and on the power of the guided mode, consequently this
effect cannot be ascribed to Faraday effect or to any other linear,
i.e. power-independent, magneto-optical phenomena. It should also
be noted that, in contrast to conventional magneto-optical effects,
in the plasmonic He-McKellar-Wilkens effect the mechanism of modulation
of interferometer transmission is not linked to material used to implement
the waveguide, instead the modulation arises as an intrinsic property
of the charge carriers that give rise to the plasmon waves. Finally
we address the scaling of the observable effect with optical power.
To remain within the quasi-particle model, we shall assume that the
asymmetry of the interferometer is small ($\cos\left(\Delta\phi_{HMW}\right)\approx1-\frac{1}{2}\Delta\phi_{HMW}^{2}$),
and thus the observable interferometer disbalance due to He-McKellar-Wilkens
effect will scale linearly with guided power.

% conclusion

In conclusion, we have proposed and analyzed a plasmonic version of
the He-McKellar-Wilkens effect. We have shown that this effect can
be observed under routinely accessible strengths of magnetic field
and power of electromagnetic radiation, using a plasmonic waveguide
interferometer. The proposed effect will allow the development of
a new generation of compact magneto-optical modulators and may be
used for tuning active plasmonic devices such as spaser \cite{Stockman03,Stockman08,LasingSpaser08}.
\begin{acknowledgments}
This study was supported by the Engineering and Physical Sciences
Research Council (grant EP/G060363/1), the Royal Society, and the
Singapore Ministry of Education {[}Grant MOE2011-T3-1-005{]}. Authors
gratefully acknowledge enlightening discussions with Prof. M. Stockman
from Georgia State University (USA). 
\end{acknowledgments}


\end{document}